\PassOptionsToPackage{hyphens}{url}
\documentclass[aps,floatfix,prx,superscriptaddress,reprint,10pt,preprintnumbers]{revtex4-2}

\input{setting.sty}

\begin{document}

\newpage

\title{Advantages of Co\nobreakdash-locating Quantum\nobreakdash-HPC Platforms:\\A Survey for Near\nobreakdash-Future Industrial Applications}

\author{Daigo Honda}
\email{dhonda@lineainc.co.jp}
\affiliation{Linea Co.,Ltd., 6-4-10 Akasaka, Minato-ku, Tokyo 107-0052, Japan}

\author{Yuta Nishiyama}
\affiliation{Linea Co.,Ltd., 6-4-10 Akasaka, Minato-ku, Tokyo 107-0052, Japan}

\author{Junya Ishikawa}
\affiliation{JSOL Corporation, Kudan-kaikan Terace 1-6-5, Kudanminami, Chiyoda-ku, Tokyo 102-0074, Japan}

\author{Kenichi Matsuzaki}
\affiliation{RIKEN SUURI Corporation, Kudan-kaikan Terace 1-6-5, Kudanminami, Chiyoda-ku, Tokyo 102-0074, Japan}

\author{Satoshi Miyata}
\affiliation{SoftBank Corp., 1-7-1 Kaigan, Minato-ku, Tokyo 105-7529, Japan}

\author{Tadahiro Chujo}
\affiliation{SoftBank Corp., 1-7-1 Kaigan, Minato-ku, Tokyo 105-7529, Japan}

\author{Yasuhisa Yamamoto}
\affiliation{SoftBank Corp., 1-7-1 Kaigan, Minato-ku, Tokyo 105-7529, Japan}

\author{Masahiko Kiminami}
\affiliation{SoftBank Corp., 1-7-1 Kaigan, Minato-ku, Tokyo 105-7529, Japan}

\author{Taro Kato}
\affiliation{SoftBank Corp., 1-7-1 Kaigan, Minato-ku, Tokyo 105-7529, Japan}

\author{Jun Towada}
\affiliation{SoftBank Corp., 1-7-1 Kaigan, Minato-ku, Tokyo 105-7529, Japan}

\author{Naoki Yoshioka}
\affiliation{RIKEN Center for Computational Science, 2-1 Hirosawa, Wako, Saitama 351-0198, Japan}

\author{Naoto Aoki}
\affiliation{RIKEN Center for Computational Science, 2-1 Hirosawa, Wako, Saitama 351-0198, Japan}

\author{Nobuyasu Ito}
\affiliation{RIKEN Center for Computational Science, 7-1-26 Minatojima-minami-machi, Chuo-ku, Kobe, Hyogo 650-0047, Japan}


\begin{abstract}
    We conducted a systematic survey of emerging quantum-HPC platforms, which integrate quantum computers and High\nobreakdash-Performance Computing (HPC) systems through co\nobreakdash-location.
    Currently, it remains unclear whether such platforms provide tangible benefits for near\nobreakdash-future industrial applications.
    To address this, we examined the impact of co\nobreakdash-location on latency reduction, bandwidth enhancement, and advanced job scheduling.
    Additionally, we assessed how HPC\nobreakdash-level capabilities could enhance hybrid algorithm performance, support large\nobreakdash-scale error mitigation, and facilitate complex quantum circuit partitioning and optimization.
    Our findings demonstrate that co\nobreakdash-locating quantum and HPC systems can yield measurable improvements in overall hybrid job throughput.
    We also observe that large\nobreakdash-scale real\nobreakdash-world problems can require HPC\nobreakdash-level computational resources for executing hybrid algorithms.
\end{abstract}

\maketitle

\section{Introduction}
\label{sec:intro}
In recent years, quantum computing has garnered significant attention as a disruptive technology with the potential to solve certain classes of problems more efficiently than conventional (classical) computers~\cite{nielsen2010quantum, montanaro2016quantum}.
However, current quantum computers remain in the Noisy Intermediate\nobreakdash-Scale Quantum (NISQ) era~\cite{preskill2018quantum}, which implies that quantum hardware possesses a limited number of qubits and is susceptible to noise\nobreakdash-induced errors.
To overcome these limitations, quantum\nobreakdash-classical hybrid algorithms offload certain computational tasks (e.g., parameter updates as well as pre\nobreakdash- and post\nobreakdash-processing tasks) to conventional computers~\cite{cerezo2021variational, bharti2022noisy}.
As the number of qubits increases or hybrid algorithms handle more practically sized problems, larger conventional computing resources are increasingly required.
This necessity has led to the proposal of ``Quantum\nobreakdash-Centric Supercomputing,'' an approach utilizing High\nobreakdash-Performance Computing (HPC) systems as the conventional computing resource~\cite{schneider2024, alexeev2024quantum}.
Accordingly, placing quantum computers in close physical proximity to HPC systems and integrating them into a unified quantum\nobreakdash-HPC platform is now regarded as a global trend, though it is still in the early stages of adoption
~\cite{
    cesga2023, 
    cleveland2023, 
    riken2023, 
    jhpc, 
    aist2024, 
    hpcqs2024, 
    innsbruck2024, 
    it4innovations2024, 
    juelich2024, 
    iqm2024, 
    brilliance2024, 
    psnc2024, 
    eurohpcju2024, 
    eurohpcju2025, 
    pawsey2025}. 

Since the foundational study~\cite{britt2017high}, numerous studies have explored the integration of quantum computers and HPC systems.
However, systematic investigations into the practical aspects of the quantum\nobreakdash-HPC platform are still emerging,
especially regarding the communication and scheduling benefits arising from co\nobreakdash-location and the advantages provided by the extensive computing resources of HPC systems.
To address this gap, we review key academic publications and supplement them with analytical insights where comprehensive data is not yet available.
Through this survey, we highlight the current state of research and propose directions for future investigation.

In particular, we clarify why communication and scheduling are critical when executing quantum\nobreakdash-classical hybrid algorithms.
These considerations are motivated by the frequent back\nobreakdash-and\nobreakdash-forth data exchanges inherent in hybrid algorithms.
Quantum computers perform specific quantum operations, while HPC nodes handle classical\nobreakdash-side tasks such as parameter optimization~\cite{cerezo2021variational, bharti2022noisy},
error mitigation~\cite{cai2023quantum}, or quantum circuit pre\nobreakdash-processing~\cite{bravyi2016trading, peng2020simulating, ge2024quantum}.
Therefore, reducing latency, increasing bandwidth, and minimizing job queueing time are crucial for achieving performance improvements in the throughput of quantum\nobreakdash-classical hybrid job execution.

This survey is structured around two major research questions. First, we examine whether the co\nobreakdash-location of quantum and conventional computers connected via high\nobreakdash-speed networks provides tangible advantages.
This encompasses three key factors that could deliver advantages: latency reduction, increased bandwidth, and advanced scheduling strategies.
Second, we examine whether leveraging a full\nobreakdash-fledged HPC system, rather than a standard conventional computer, results in additional benefits.
In particular, we focus on three key capabilities facilitated by HPC systems: quantum\nobreakdash-classical hybrid algorithm execution, error mitigation methods through extensive data post\nobreakdash-processing, and quantum circuit partitioning and optimization procedures.

In this survey, we primarily target near\nobreakdash-future industrial use cases, envisioned to mature by 2030, that involve the execution of quantum\nobreakdash-classical hybrid algorithms on NISQ computers.
We focus on use cases closely related to industrial applications, such as quantum chemistry calculations (e.g., the Variational Quantum Eigensolver, VQE~\cite{peruzzo2014variational})
and quantum machine learning techniques (e.g., quantum circuit learning~\cite{mitarai2018quantum}, quantum neural networks~\cite{farhi2018classification}).
While these applications illustrate promising directions, the primary contribution of this survey is to evaluate their computational feasibility on quantum\nobreakdash-HPC platforms from a computational science perspective.

Through this survey, we identified evidence from existing literature that positively addresses the two research questions.
Our findings suggest that the co\nobreakdash-location of quantum and conventional computers within an integrated platform can deliver measurable enhancements in the overall throughput of hybrid algorithm execution.
We also observe that addressing large\nobreakdash-scale real\nobreakdash-world problems requires handling larger quantum circuits, leading to increasingly heavy classical\nobreakdash-side computational loads in both pre\nobreakdash-processing and post\nobreakdash-processing stages.
Consequently, this raises the potential necessity for HPC\nobreakdash-level computational resources.
Indeed, we found a notable example that explicitly employed HPC resources.
However, the existing evidence primarily addresses isolated aspects rather than a comprehensive view.
We anticipate that future research will conduct comprehensive evaluations from an integrated, holistic perspective to unlock new computational regimes and accelerate practical, large\nobreakdash-scale quantum\nobreakdash-classical hybrid computation.

The remainder of this paper is organized as follows.
Section~\ref{sec:overview} provides an overview of our analytical framework, introducing an envisioned quantum\nobreakdash-HPC platform.
Section~\ref{sec:quantum-classical-colocation} addresses the first research question by discussing the communication and scheduling benefits achieved through the co\nobreakdash-location of quantum and conventional computers.
Section~\ref{sec:hpc-resource} addresses the second research question by exploring how HPC resources could potentially enhance the performance of quantum\nobreakdash-classical hybrid computing.
Finally, Section~\ref{sec:conclusion} summarizes our findings and suggests future research directions.

\section{Overview of Analytical Framework}
\label{sec:overview}

\begin{figure*}[tb]
    \centering
    \includegraphics[width=\textwidth]{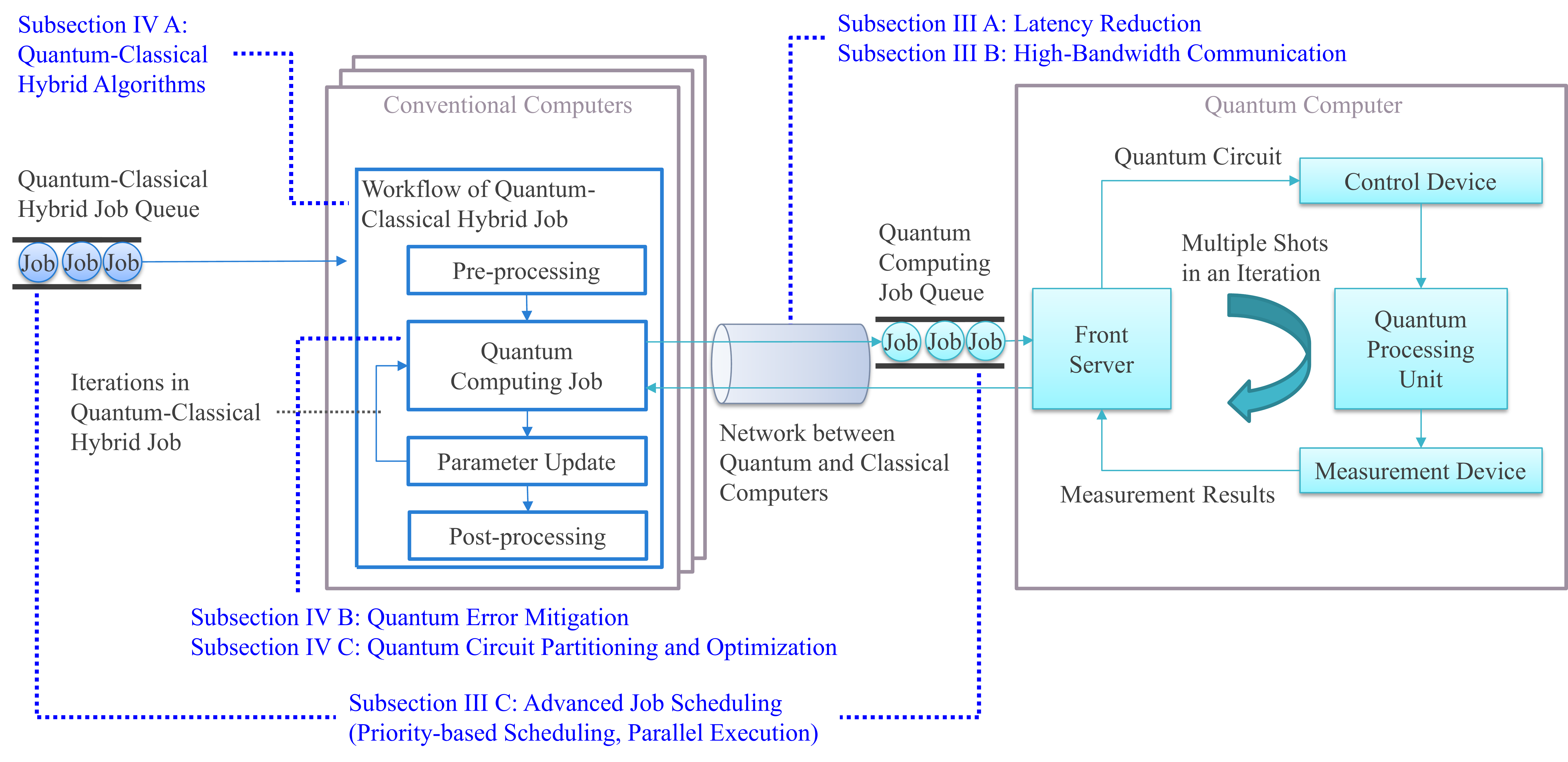}
    \caption{An overview of our analytical framework for evaluating the advantages of co-locating quantum-HPC platforms.}
    \label{fig:overview_qc_hpc}
\end{figure*}

In this section, we provide an overview of our analytical framework by presenting the envisioned quantum\nobreakdash-HPC platform's system architecture
and describing the workflow for quantum\nobreakdash-classical hybrid jobs.
We also specify which framework components address each research question and refer to Figure~\ref{fig:overview_qc_hpc} for a visual overview.
Further details are provided in the following paragraphs.

First, we describe the system architecture and workflows by contrasting two scenarios: one in which only quantum computers are used, and another in which quantum and conventional computers are integrated.
When users execute standalone quantum computing jobs, they typically submit them from conventional computers via the internet to quantum computers, which then queue and schedule them.
Requests are first received by the quantum computer's front server, and the control device operates the Quantum Processing Unit (QPU) to run the quantum circuits.
The measurement device then reads out the qubits and sends the results back to the front server.
Finally, the results are returned to the respective conventional computers or fetched once job completion is detected.

In contrast to the standalone scenario, quantum\nobreakdash-classical hybrid jobs typically involve multiple iterations in their workflows.
After pre\nobreakdash-processing, the workflow repeats iterations of quantum computing job executions and parameter updates on conventional computers.
Once the necessary iterations are completed, post\nobreakdash-processing is conducted, concluding the workflow of the hybrid job.

Each iteration incurs communication delays as data moves between quantum and conventional computers.
Moreover, if each quantum computing job within a hybrid job is scheduled in the same manner as unrelated jobs,
the hybrid job repeatedly experiences waiting times due to scheduling delays.
Such an architecture is not ideal for hybrid jobs, which require tightly coupled integration and rapid communication between quantum and conventional computers.

The first research question posed in the Introduction is about the advantage of the co\nobreakdash-location of quantum and conventional computers.
Our hypothesis is that the aforementioned communication and scheduling inefficiencies can be resolved by co\nobreakdash-location.
Specifically, we explore the potential of a unified platform where quantum and conventional computers are interconnected through a low\nobreakdash-latency and high\nobreakdash-bandwidth network and managed via coordinated scheduling.
This setup allows quantum\nobreakdash-classical hybrid jobs to be executed as unified workflows rather than as a series of disconnected tasks, thereby potentially enhancing overall execution efficiency.

From an operational perspective, achieving high job throughput is often more important than minimizing the runtime of a single job.
In multi\nobreakdash-user scenarios, numerous quantum\nobreakdash-classical hybrid jobs may be queued for execution simultaneously.
By reducing communication and scheduling delays through co\nobreakdash-location, each job experiences a shorter turnaround time,
thereby increasing the total number of jobs processed in a given period.
Consequently, this survey pays particular attention to the impact on the throughput of hybrid jobs, rather than the performance of individual jobs.

The second research question posed in the Introduction is about the benefit of integrating HPC as conventional computing resources utilized in executing quantum\nobreakdash-classical hybrid algorithms.
Here we investigate hybrid algorithms in fields such as quantum chemistry and quantum machine learning, especially in the context of NISQ computers.
Within the workflows of these algorithms, conventional computing resources are utilized for tasks such as parameter updates during quantum\nobreakdash-classical iterations,
as well as pre\nobreakdash-/post\nobreakdash-processing tasks, including error mitigation and quantum circuit partitioning and optimization.

We hypothesize that these algorithms require increasingly substantial conventional computing resources as they are applied to large\nobreakdash-scale, real\nobreakdash-world problems.
However, it remains to be investigated whether these growing demands truly necessitate HPC resources to ensure overall system efficiency.

\section{Communication and Scheduling Benefits of Co-locating Quantum and Conventional Computers}
\label{sec:quantum-classical-colocation}

In this section, we address the first research question posed in the Introduction:
whether the co\nobreakdash-location of quantum and conventional computers connected via high\nobreakdash-speed networks provides tangible advantages.
A comprehensive overview of the analytical framework and the underlying hypotheses is provided in Section~\ref{sec:overview}.

To explore these potential advantages, the following subsections discuss the existing literature and experimental results that support three primary benefits of the co\nobreakdash-location.
In Subsection~\ref{subsec:latency_reduction}, we investigate how latency reduction can improve job throughput.
Next, in Subsection~\ref{subsec:high_bandwidth}, we examine the benefits of high\nobreakdash-bandwidth communication gained from efficiently exchanging large volumes of intermediate data between quantum and conventional computers.
Finally, in Subsection~\ref{subsec:job_scheduling}, we discuss how advanced scheduling across quantum and conventional computers reduces queueing time and enhances job throughput.

When examining the communication overhead between quantum and conventional computers,
we explicitly treat latency and data transmission time, defined as data volume divided by bandwidth, as independent parameters.
For clarity, all references to ``latency'' herein refer specifically to the one\nobreakdash-way latency between the quantum and conventional computers, unless otherwise noted.

We also note that our survey specifically targets near\nobreakdash-term applications in fields such as quantum chemistry and quantum machine learning, both of which utilize NISQ computers.

\subsection{Latency Reduction}
\label{subsec:latency_reduction}

This subsection investigates the potential advantages of reducing latency between quantum and conventional computers.
We focus on situations in which the volume of data transferred is relatively small,
making latency the dominant factor in communication overhead.
Although an increase in data volume inevitably highlights the importance of bandwidth,
we discuss such scenarios in Subsection~\ref{subsec:high_bandwidth}.

A key characteristic of quantum\nobreakdash-classical hybrid algorithms is their iterative structure, in which quantum\nobreakdash-side computations alternate with classical\nobreakdash-side computations.
Each iteration thus incurs a round\nobreakdash-trip communication cost that accumulates over multiple iterations.
In current deployments, quantum computers are often placed far from conventional computers, leading to non\nobreakdash-negligible latency.
This latency can adversely impact the efficiency of hybrid computations by increasing total runtimes.

Against this backdrop, we hypothesize that co\nobreakdash-locating quantum and conventional computers and connecting them via high\nobreakdash-speed networks
can reduce latency and thereby shorten the execution time of each quantum\nobreakdash-classical hybrid job.
This reduction, in turn, collectively increases the throughput of hybrid jobs.

In our survey, we did not find any measured data on latency and throughput in existing platforms integrating quantum and conventional computers;
however, we identified a study that used a model\nobreakdash-based analysis to discuss the relationship between latency and throughput~\cite{farooqi2023exploring}.
We applied our own proposed latency scenarios to this model to quantify the potential improvements in throughput achievable by reducing latency.

\subsubsection{Latency in Co-located vs. Remotely Located Configurations}
We examine the latency that occurs when the quantum computer is co\nobreakdash-located or remotely located relative to the conventional computers.
In a co\nobreakdash-located scenario, the quantum and conventional computers communicate over a Local Area Network (LAN).
In contrast, a remotely located scenario typically relies on a Wide Area Network (WAN) connection.

To establish realistic latency scenarios, we refer to the JHPC\nobreakdash-quantum project~\cite{riken2023, jhpc} as a case study.
The JHPC\nobreakdash-quantum project has deployed quantum computers in two distinct geographical configurations relative to the supercomputer Fugaku in Kobe, Japan.
We derive scenarios for latency from these configurations rather than relying on actual measurements.

In one configuration, an IBM superconducting quantum computer is co\nobreakdash-located with Fugaku in the same building.
This corresponds to the co\nobreakdash-located scenario. In this scenario, the effective latency is
largely determined by the networking technology and the specific construction of the local environment, rather than distance alone.
High\nobreakdash-performance networks such as InfiniBand are commonly employed in HPC clusters and data centers
to achieve low latency and high bandwidth over relatively short distances (on the order of up to 100 meters).
Although dependent on the specific configuration, for reference,
the report~\cite{fsus2023} states that the end\nobreakdash-to\nobreakdash-end latency at the application layer
within the same cluster is around 2 \si{\us} in InfiniBand environments.
Based on the above, we assume a latency of 2 \si{\us} in the co\nobreakdash-located scenario.

In another configuration, a Quantinuum trapped\nobreakdash-ion quantum computer is installed at RIKEN's Wako campus,
which is approximately 400 \si{\km} away from Kobe in a straight line.
When quantum and conventional computers are geographically separated, the propagation delay in optical fibers becomes dominant.
Light travels through fiber at roughly 200,000 \si{\km/\s}, implying a 0.5 \si{\ms} one\nobreakdash-way delay per 100 km under ideal conditions.
Additional latency accrues from intermediary network devices such as routers and switches, making precise estimates challenging.
Nevertheless, the distance\nobreakdash-based propagation delay itself often dominates when the quantum and conventional computers are located
hundreds or thousands of kilometers apart.
Based on the above, we assume a latency of 2 \si{\ms} in the remotely located scenario.

For comparison, we further consider a hypothetical quantum computer situated in North America, about 10,000 km away,
which would result in roughly 50 \si{\ms} of one\nobreakdash-way latency.

\subsubsection{Latency and Throughput Modeling}
We first describe the model used in our analysis, which is described in~\cite{farooqi2023exploring}.
This model evaluates how the latency between quantum and conventional computers affects
the number of hybrid jobs that can be completed per hour (i.e., throughput).
For simplicity, it assumes many identical hybrid jobs are executed,
and defines throughput according to the following equation:

\begin{equation}
    \text{Throughput} = \frac{A}{2 \times 10^x + B}, \label{eq:throughput_model}
\end{equation}
where
\begin{align}
    x &= \log_{10}(\text{latency}), \label{eq:definition_x}\\
    A &= \frac{3600}{\text{number of iterations}}, \\
    B &= \text{overhead} + \text{classical-side computation time} \\
    +& (\text{quantum circuit execution time} \times \text{number of shots}). \nonumber
\end{align}
In this context, the overhead refers to the sum of (i) the latency within the conventional compute nodes,
(ii) the control electronics reprogramming time and (iii) the qubit readout time,
excluding the latency between the quantum and conventional computers.
All time values in the model are converted into seconds for consistency.

The parameter sets used in our analysis are summarized in Table~\ref{tab:parameter_set}.
These sets define different job configurations to explore how various factors affect throughput.
\begin{itemize}
    \item Pattern 1 corresponds to the highest\nobreakdash-throughput scenario described in~\cite{farooqi2023exploring}.
    \item Pattern 2 assumes a longer quantum circuit execution time, reflecting the fact that actual runtimes can vary depending on the quantum hardware (e.g., superconducting vs. trapped\nobreakdash-ion systems).
    \item Pattern 2' instead increases the number of shots, capturing scenarios where higher sampling is required.
  \end{itemize}
Note that Pattern 2 and Pattern 2' share the same values for the model parameters \(A\) and \(B\).
This indicates that increasing the number of shots influences throughput similarly to an increase in quantum circuit execution time.

\begin{table*}[tb]
    \centering
    \caption{Parameter sets used in our model-based analysis.}
    \label{tab:parameter_set}
    \begin{tabular}{lrrr}
        \hline
        \textbf{Parameter} & \textbf{Pattern 1} & \textbf{Pattern 2} & \textbf{Pattern 2'}\\
        \hline
        number of iterations & 100 & 100 & 100 \\
        overhead (\si{\ms}) & 10.501 & 10.501 & 10.501\\
        classical\nobreakdash-side computation time (\si{\ms}) & 1 & 1 & 1 \\
        quantum circuit execution time (\si{\ms}) & 0.05 & 0.5 & 0.05\\
        number of shots & 1000 & 1000 & 10,000 \\
        \hline
        model parameter \(A\) (\si{\s})& 36 & 36 & 36\\
        model parameter \(B\) (\si{\s})& 0.061501 & 0.511501 & 0.511501\\
        \hline
    \end{tabular}
\end{table*}

\begin{figure*}[tb]
    \centering
    \includegraphics[width=\textwidth]{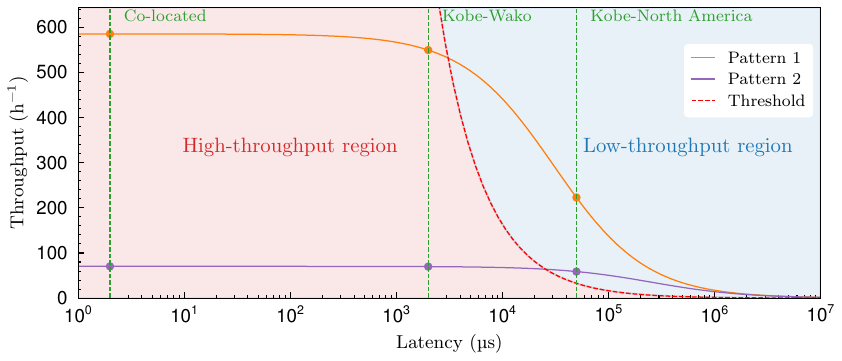}
    \caption{Relationship between latency (log scale, in microseconds) and throughput (jobs per hour) for two parameter sets.
        The orange and purple curves represent Patterns 1 and 2, respectively.
        The red dashed line indicates the threshold trajectory, dividing the high-throughput region (shaded red) and the low-throughput region (shaded blue).
        The vertical green dashed lines represent latency values for the co-located scenario (2 \si{\us}) and remotely located scenarios between Kobe and Wako (2 ms), and Kobe and North America (50 ms).}
    \label{fig:throughput_latency}
\end{figure*}

Figure~\ref{fig:throughput_latency} shows the throughput values calculated by this model for different latency values \(x\).
An important finding from this figure is the existence of a threshold, beyond which even small increases in latency cause a significant drop in throughput.
In the high\nobreakdash-throughput region, where latency is well below the threshold, the throughput remains relatively stable
even if the latency varies.
However, once in the low\nobreakdash-throughput region, incremental increases in latency can dramatically reduce the achievable throughput.

While this threshold value is not defined explicitly in Reference~\cite{farooqi2023exploring}, one reasonable approach is to define
the threshold using the following equation:
\begin{equation}
    2 \times 10^x = \frac{B}{10} \label{eq:thresholds}
\end{equation}
beyond which the slope of the throughput curve begins to steepen.
At this threshold, the throughput reaches approximately 90\% of its theoretical maximum \( A/B \) (i.e., the limit as latency goes to zero).
This definition implies that the order of the parameter \(B\),
which is the time components other than the latency between the quantum and conventional computers,
essentially determines the threshold.
In Figure~\ref{fig:throughput_latency}, we plot the trajectory of the threshold and the throughput value at that threshold with varying parameter \(B\),
shown by the red dashed line.
The left side of this trajectory corresponds to the high\nobreakdash-throughput region, while the right side corresponds to the low\nobreakdash-throughput region.

Applying this model to the latency scenarios introduced earlier provides insight into
whether a particular configuration lies in the high\nobreakdash-throughput or low\nobreakdash-throughput region.
\paragraph*{Co-located scenario}
With a latency of 2 \si{\us}, this scenario is in the high\nobreakdash-throughput regime for both Pattern 1 and Pattern 2, defined in Table~\ref{tab:parameter_set}.
Even if the threshold shifts due to parameter variations, 2 \si{\us} remains far below any realistic threshold.
\paragraph*{Remotely located scenario}
For both Pattern 1 and Pattern 2, a latency of 2 \si{\ms} (e.g., between Kobe and Wako) remains in the high\nobreakdash-throughput region,
while a latency of 50 \si{\ms} (e.g., between Kobe and North America) falls into the low\nobreakdash-throughput region.
However, the threshold shifts closer to 50 \si{\ms} in Pattern 2 due to the longer quantum circuit execution time,
and if the threshold shifted even higher, the 50 \si{\ms} latency would eventually enter the high\nobreakdash-throughput region.
Thus, as the threshold increases, the relative advantage gained by latency reduction diminishes.
Detailed throughput values and corresponding improvements gained by co\nobreakdash-location for each scenario are summarized in Table~\ref{tab:throughput_comparison}.

\begin{table*}[tb]
    \centering
    \caption{Throughput comparison and relative improvement rates under different latency scenarios.}
    \label{tab:throughput_comparison}
    \begin{tabular}{lrrr}
        \hline
        \textbf{Scenario} & \textbf{Latency} & \textbf{Throughput} & \textbf{Improvement by co\nobreakdash-location} \\
        \hline
        \textbf{Pattern 1} & & & \\
        \quad Co\nobreakdash-located & 2 \si{\us} & 585.3 & -- \\
        \quad Kobe--Wako & 2 \si{\ms} & 549.6 & 6.5\% \\
        \quad Kobe--North America & 50 \si{\ms} & 222.9 & 162.6\% \\
        \hline
        \textbf{Pattern 2} & & & \\
        \quad Co\nobreakdash-located & 2 \si{\us} & 70.4 & -- \\
        \quad Kobe--Wako & 2 \si{\ms} & 69.8 & 0.9\% \\
        \quad Kobe--North America & 50 \si{\ms} & 58.9 & 19.5\% \\
        \hline
    \end{tabular}
\end{table*}

These outcomes emphasize that the benefit of reducing latency is highly dependent on the threshold.
The threshold depends not only on the quantum circuit execution time but also on the number of shots.
For instance, in Pattern 2', increasing the number of shots leads the model to behave similarly to Pattern 2,
indicating that both a longer quantum circuit execution time and a larger number of shots raise the threshold in the same way.
As a result, as the number of shots increases, the threshold grows, thereby diminishing the benefit of latency reduction.

Thus, while hardware choice and proximity certainly affect the throughput,
the characteristics of hybrid jobs, such as the number of shots, can also shift the threshold
and reduce the benefits of co\nobreakdash-locating quantum and conventional computers.
In particular, the number of shots is a user\nobreakdash-defined parameter and can vary significantly,
even when running the same quantum circuit for the same purpose.

\subsubsection{Case Study: Number of Shots}
Our model\nobreakdash-based analysis confirmed that increasing the number of shots contributes to the increase in the latency threshold,
potentially reducing the overall benefit of lower latency.
Here, we examine whether the number of shots might exceed the initially assumed value of 1000.
To do so, we reviewed research from both quantum chemistry and quantum machine learning that discusses variations in the number of shots.

As an example from quantum chemistry, Reference~\cite{mihalikova2022cost} focused on how varying the number of shots in a single iteration
of the VQE affected computational accuracy.
In their benchmark study, calculating the ground\nobreakdash-state energy of the hydrogen atom, the authors ran VQE on a Qiskit QASM simulator
with 512, 1024, 4096, and 8192 shots per iteration.
Their results show that increasing the number of shots improves accuracy.

In contrast to the previous study's focus on increasing shots, Reference~\cite{zhu2024optimizing} explored strategies to control the increase of the total number of shots used in VQE
by imposing a shot budget and employing an optimal allocation strategy.
They used the ground\nobreakdash-state energy calculation of LiH as a benchmark.
Specifically, they examined how to distribute 27,000 shots among the 27 Pauli terms in the LiH Hamiltonian.
Under a uniform assignment, each term would receive 1000 shots, but the authors proposed a ``Variance\nobreakdash-Preserved Shot Reduction'' method
that numerically demonstrated a reduction in the total number of shots needed to achieve VQE convergence.

As an example from quantum machine learning, Reference~\cite{delgado2022unsupervised} examined the impact of varying the number of shots in quantum circuit Born machine training.
The authors studied a generative model for reproducing the kinematic distributions of particle jets in high\nobreakdash-energy collider experiments.
They ran a Qiskit QASM simulator with 8192 and 20,000 shots per iteration, and they observed that increasing the number of shots reduced the Jensen\nobreakdash-Shannon divergence during training.

Another study~\cite{miroszewski2024search} provided a theoretical estimate of the number of shots required to evaluate quantum kernel matrices.
For instance, based on their specific settings, it was estimated that evaluating a single element of the quantum kernel matrix with a ZZ\nobreakdash-feature map of around 20 qubits would require on the order of several million shots.
A phenomenon known as exponential concentration~\cite{huang2021power,kubler2021inductive,thanasilp2024exponential} explains why evaluating quantum kernels requires such a large number of shots.
Exponential concentration implies that distinguishing small differences in quantum kernel values requires exponentially more shots as the number of qubits increases.
Since the IBM Quantum platform limits a single job to five million circuit executions~\cite{ibm_joblimits},
further exponential increases would pose practical constraints for actual runtime experiments.

Although some works~\cite{mihalikova2022cost,delgado2022unsupervised,miroszewski2024search} suggest that the number of shots could grow substantially beyond 1000, others~\cite{zhu2024optimizing} emphasize shot\nobreakdash-reduction strategies.
Whether the number of shots will continue to increase remains uncertain, as the optimum setting depends on factors such as application context and target accuracy.
Therefore, it cannot be definitively stated that latency reduction will become less beneficial for all hybrid jobs in the future.

\subsubsection{Conclusion}
We have investigated how co\nobreakdash-locating quantum and conventional computers can benefit hybrid jobs by reducing latency and thus improving job throughput.
Because actual measurements of latency and throughput in such environments were not available, our evaluation relied on theoretical modeling and scenario\nobreakdash-based analyses.

Our model\nobreakdash-based analysis supports the notion that co\nobreakdash-locating quantum and conventional computers, compared to remotely located configurations, can yield higher job throughput through latency reduction.
However, some hybrid jobs experience smaller gains depending on the quantum hardware choice (such as superconducting or trapped\nobreakdash-ion systems) or a large number of shots per iteration.

Even if certain jobs experience smaller gains, in practice actual platforms execute a variety of jobs, each yielding differing gains, rather than repeatedly running identical jobs.
Therefore, the advantage of co\nobreakdash-location still exists when evaluating overall throughput on the hybrid computing platform.

\subsection{High-Bandwidth Communication}
\label{subsec:high_bandwidth}

This subsection investigates the advantages of increasing network bandwidth enabling large data communication between quantum and conventional computers.
Our hypothesis for the advantage of high\nobreakdash-bandwidth communication involves error mitigation in quantum computations, which potentially causes large measurement data.
In the NISQ era, error mitigation strategies are crucial due to non\nobreakdash-negligible noise that inevitably arises during quantum computation.
Without applying error mitigation, the amount of data generated per shot is proportional to the order of the number of qubits involved,
typically hundreds of bits at most in current NISQ computers.
Consequently, the total data from roughly 1000 shots typically remains within tens to hundreds of kilobytes, from simple calculation considering only the bit string of measurement data~\footnote
{
    Assuming that NISQ computers possess between 100 and 1000 qubits, we estimated the data volume of bit strings generated by 1000 shots.
    100 bits \(\times\) 1000 shots \(=\) 12,500 bytes.
    1000 bits \(\times\) 1000 shots \(=\) 125,000 bytes.
}.

However, the implementation of error mitigation techniques requires additional measurements to estimate the noise\nobreakdash-free results accurately.
This process can involve executing the desired quantum circuit multiple times or utilizing supplementary quantum circuits specifically designed for error characterization.
Increasing the number of quantum circuit executions naturally leads to growth in measurement data volume that must be transmitted to conventional computers.
It is known that as the number of qubits and the quantum circuit depth grow, the required number of circuit executions and the size of measurement data increase exponentially~\cite{takagi2023universal,tsubouchi2023universal}.

We conducted our investigation to test this hypothesis.
In the context of this subsection, the prefixes k, M, and G strictly follow the SI unit convention (k \(= 10^3\), M \(= 10^6\), G \(= 10^9\)).

\subsubsection{Bandwidth in Co-located vs. Remotely Located Configurations}
When quantum computers are co\nobreakdash-located with conventional computers, communication typically occurs via a LAN.
In contrast, when they are remotely located, they rely on a WAN.

In co\nobreakdash-located scenarios, high\nobreakdash-bandwidth data communication can be achieved using specialized networking technologies such as InfiniBand.
For instance, the InfiniBand Next Data Rate (NDR) specification provides a bandwidth of up to 400 Gbps~\cite{lu2022survey}.

On the other hand, remotely located scenarios that rely on the public WAN connections typically experience significantly lower and
less stable effective throughput compared to LAN configurations.
Using dedicated WAN infrastructure, however, can improve both bandwidth and reliability.
One such dedicated WAN infrastructure is Japan's scientific information network, SINET6, which can deliver up to 400 Gbps between remote sites~\cite{sinet6}.
Nevertheless, inherent issues such as packet retransmissions, congestion control, and the shared nature of a WAN can still degrade performance on dedicated WANs, including SINET6~\cite{peterson2012computer}.
As a result, the actual usable bandwidth may decrease, and the stability and bandwidth levels cannot be guaranteed to match those typically achievable through InfiniBand\nobreakdash-based LAN connections.

\subsubsection{Impact of Data Transmission Time on Throughput}

Before discussing the increase in data volumes, it is essential to revisit and refine the analysis presented in Subsection~\ref{subsec:latency_reduction}
to consider scenarios in which data transmission time is no longer negligible.
The original throughput model equation~\eqref{eq:throughput_model} has a key parameter, denoted by \(x\).
As defined in the equation~\eqref{eq:definition_x}, \(x\) is the common logarithm of the latency alone.
However, to accurately represent realistic conditions, this parameter must be revised to account for the total communication time.
Here we define total communication time as the sum of latency and data transmission time.
It is important to note that the latency referred to here is one\nobreakdash-way.
Therefore, the data transmission time must also be treated as one\nobreakdash-way.
In this subsection, we consider scenarios where only the data transmitted from a quantum computer to a conventional computer predominantly increases.
Nevertheless, for throughput modeling purposes, it is acceptable to simply use the average data volume of the round\nobreakdash-trip communication.
Therefore, the definition of \(x\) is revised as follows:
\begin{equation}
    x = \log_{10}\left(\text{latency} + \frac{1}{2} \times \frac{\text{data volume}}{\text{bandwidth}}\right),
\end{equation}
where the factor of 1/2 accounts for one\nobreakdash-way data transmission, assuming the total data volume is for a round trip.

To evaluate the potential impact of this modification, let us consider a simplified scenario based on previously discussed conditions.
As noted at the beginning of this subsection, when error mitigation is not employed, the data volume resulting from approximately 1000 shots typically ranges from tens to hundreds of kilobytes.
For simplicity, we assume a representative data volume of 100 kB.
Further assuming the co\nobreakdash-located scenario conditions with a latency of 2 \si{\us} and a bandwidth of 400 Gbps, the data transmission time computed by the above formula is 1 \si{\us}.
Consequently, the total communication time becomes 3 \si{\us}.
Under such conditions, the total communication time remains within the high\nobreakdash-throughput region, indicating negligible impact on throughput.

However, as data volumes increase or when using networks with lower bandwidth, the data transmission time grows,
potentially causing the total communication time to exceed thresholds that significantly diminish throughput.
These thresholds can be determined by applying the revised definition of \(x\) to equation~\eqref{eq:thresholds}.
Specifically, by substituting the revised \(x\) and a latency of 2 \si{\us} into equation~\eqref{eq:thresholds}, we obtain the following expression:
\begin{equation}
    2\times \left(2~\si{\us} + \frac{1}{2} \times \frac{\text{data volume}}{\text{bandwidth}}\right) = \frac{B}{10}.
\end{equation}
To maintain high throughput, \(x\) must be below the threshold.
Here, we conduct our analysis using Pattern 1 defined in Table~\ref{tab:parameter_set}, which places stricter requirements on \(x\).
Since the latency is negligible compared to parameter \(B\), the following inequality must be satisfied for the data transmission time to remain below the threshold:
\begin{equation}
   \frac{\text{data volume}}{\text{bandwidth}} \leq \frac{B}{10} \approx 0.00615~\si{\s}. \label{eq:threshold_inequality}
\end{equation}

To satisfy inequality~\eqref{eq:threshold_inequality}, for a data volume of 100 kB, a minimum bandwidth of approximately 130 Mbps is required.
Furthermore, if the data volume reaches 307.5 MB, the required bandwidth escalates to 400 Gbps.

\subsubsection{Data Volumes for Error Mitigation}
To better understand the potential volume of measurement data when applying error mitigation,
we review general theoretical results and specific empirical cases.

As a general consideration, universal cost bounds for existing error mitigation techniques indicate that the number of shots
required to achieve a certain computational accuracy increases exponentially with the circuit depth~\cite{takagi2023universal,tsubouchi2023universal}.
Furthermore, for random circuits with local noise, the required number of shots grows exponentially with the number of qubits~\cite{tsubouchi2023universal}.
This exponential growth in the number of shots directly implies an exponential increase in the volume of measurement data transmitted to conventional computers.

We further investigated representative examples of applying different error mitigation methods to estimate the practical scale of measurement data generated.
Specifically, we examined methods mitigating errors arising from gate operations and qubit measurements.

To mitigate gate operation errors, we considered the Zero Noise Extrapolation (ZNE) method.
ZNE involves running quantum circuits at various artificially increased noise levels and then extrapolating these results to estimate outcomes in the noise\nobreakdash-free limit~\cite{li2017efficient, temme2017error}.
Techniques used to artificially amplify noise include extending gate operation durations or employing redundant quantum circuits.

In Reference~\cite{kim2023scalable}, ZNE was applied to a benchmark quantum circuit simulating the quench dynamics of the two\nobreakdash-dimensional Ising model, using up to 26 qubits.
Error\nobreakdash-amplified circuits were generated at three distinct noise levels, and each circuit was executed 100,000 times.
The total number of shots was 300,000.
However, assuming that different circuits were executed as separate iterations, the number of shots per iteration was 100,000.
Considering only the bit string of measurement data, this experiment produces approximately 325 kB
(26 bits \(\times\) 100,000 shots \(=\) 325,000 bytes).

To mitigate measurement errors, we reviewed the Matrix\nobreakdash-free Measurement Mitigation (M3) method~\cite{nation2021scalable}.
Canonical measurement error mitigation methods become computationally infeasible as the number of qubits increases.
Thus, M3 is based on assumptions regarding measurement noise and computational optimizations, improving scalability for a larger number of qubits.

Reference~\cite{pokharel2024scalable} demonstrated M3 and a proposed method referred to as Iterative Bayesian Unfolding (IBU),
applied to a benchmark test measuring Greenberger-Horne-Zeilinger (GHZ) states with up to 127 qubits.
The experiment executed the circuit 100,000 times.
In this setting, the resulting measurement data is approximately 1.6 MB (127 bits \(\times\) 100,000 shots \(=\) 1,587,500 bytes).

However, this experiment showed a significant increase in the discrepancy between ideal and mitigated probability distributions as the number of qubits rose.
In the above study, the number of shots remained fixed at 100,000, likely contributing to the observed rapid degradation in error mitigation accuracy.
The numerical optimizations employed by M3 do not fundamentally alter the exponential scaling trend, as theoretically demonstrated for measurement error mitigation
(see Reference~\cite{bravyi2021mitigating} or brief summary of the reference described in the Appendix).
Thus, similar exponential growth in shots is theoretically expected for a higher number of qubits.
We suggest that achieving higher accuracy would require much more measurement data.

The two error mitigation methods discussed above address different types of errors and are not mutually exclusive.
Indeed, it is known that combining these methods can enhance overall accuracy~\cite{weaving2023benchmarking, cai2023quantum}.
However, employing multiple methods simultaneously increases the variety of quantum circuit executions required,
potentially leading to a further growth in the total number of shots, and therefore further growth in the measurement data volume.

\subsubsection{Conclusion}
We have investigated the advantages of co\nobreakdash-locating quantum and conventional computers from the viewpoint of enhanced bandwidth capabilities.
Specifically, we focused on the necessity of error mitigation techniques in the NISQ era
and highlighted the potentially significant increase in data volume associated with these methods.

Through the refinement of the analysis presented in Subsection~\ref{subsec:latency_reduction},
we estimated that when the data volume reaches 307.5 MB, maintaining the high throughput requires communication with a bandwidth of 400 Gbps.

We examined two practical examples that indicated data volumes ranging from several hundred kilobytes to several megabytes.
Given these current data volumes, the 400 Gbps bandwidth offered by technologies like InfiniBand may be more than necessary at present.

However, according to the universal cost bounds for existing error mitigation techniques~\cite{takagi2023universal,tsubouchi2023universal},
as the number of qubits and circuit depths increase, the volume of data required for effective error mitigation will continue to grow exponentially.
This data growth will be further accelerated by the rapid increase in the number of qubits projected in various quantum hardware development roadmaps.

Given this exponential growth of the data volume, the margin before reaching about 300 MB may not remain large for long.
Therefore, even within the near\nobreakdash-term timeframe up to 2030,
we anticipate that co\nobreakdash-locating quantum and conventional computers could offer significant benefits by establishing stable, high\nobreakdash-bandwidth communication infrastructures.

\subsection{Advanced Job Scheduling}
\label{subsec:job_scheduling}

This subsection investigates the potential advantages of advanced job scheduling strategies
to improve the overall throughput of quantum\nobreakdash-classical hybrid computing platforms.

\begin{figure*}[tbh]
    \centering
    \includegraphics[width=\textwidth]{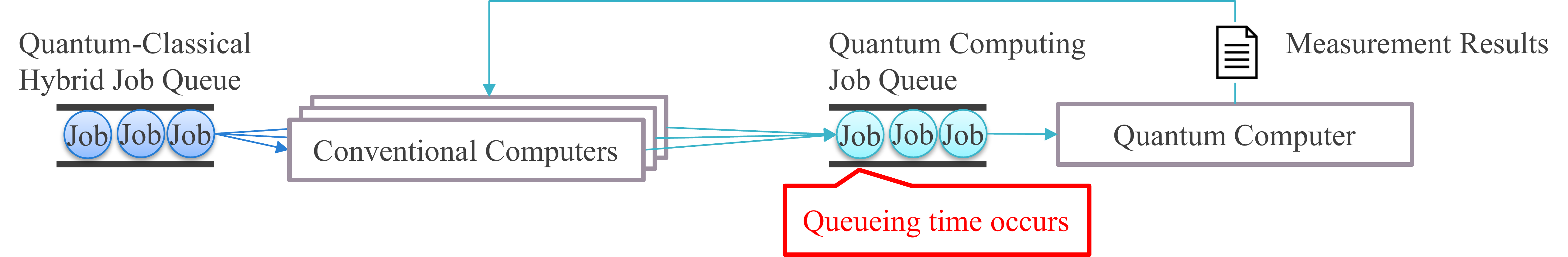}
    \caption{Quantum computing job queue and associated queueing time in quantum-classical hybrid job workflows.}
    \label{fig:queueing}
\end{figure*}

Quantum computers are currently scarce resources typically shared among multiple users,
with workloads originating from both standalone quantum computing jobs and quantum\nobreakdash-classical hybrid jobs.
Jobs submitted to quantum computers are queued, leading to queueing periods before execution.
Throughout this subsection, ``queueing time'' refers to the duration each job remains in the quantum computing job queue (see Figure~\ref{fig:queueing}).

Quantum\nobreakdash-classical hybrid jobs inherently involve numerous iterations where computational tasks alternate between conventional and quantum computers.
Under a simple First\nobreakdash-In\nobreakdash-First\nobreakdash-Out (FIFO) queueing approach, each iteration of a hybrid job returns to the end of the quantum computing job queue,
causing repeated and potentially long queueing times.
If the queueing time is long compared to the total computation and overhead time of each hybrid job (including both quantum and classical sides),
the overall execution time can be severely impacted, with the queueing time dominating the total runtime.
This prolonged queueing time not only extends execution durations but also leads to inefficient utilization of conventional computing resources,
as conventional computers remain idle during queue waiting periods.

We hypothesize that prioritizing quantum computing jobs associated with hybrid jobs can significantly improve execution efficiency.
To examine this hypothesis, we conducted a series of investigations focusing on scheduling strategies.
Prioritized execution could reduce runtime for individual hybrid jobs, thus minimizing the idle periods for conventional computing resources.
Additionally, since reducing average queueing time could enhance overall throughput when managing multiple concurrent hybrid jobs,
we explored further scheduling strategies beyond simple prioritized execution.
One such candidate is a parallel execution strategy, which enables simultaneous execution of multiple quantum computing jobs on a single QPU~\cite{das2019case}.

\subsubsection{Analysis of Queueing Times}
We examine queueing times typically observed in quantum computing job queues.
We then discuss how these queueing delays affect the total execution time of a quantum\nobreakdash-classical hybrid job.

In Reference~\cite{ravi2021quantum}, the authors analyzed queueing times for quantum computing jobs executed
on the IBM Quantum platform over two years up to April 2021.
According to this analysis, about 20\% of jobs experienced queueing times of one minute or less,
whereas the remaining 80\% exceeded one minute.
The median queueing time was approximately one hour, and more than 30\% of jobs encountered queueing times exceeding two hours.

To further quantify the impact of these queueing times on overall execution,
we now perform a calculation based on the hybrid job scenario defined in Subsection~\ref{subsec:latency_reduction}.
Specifically, we use Pattern 1 defined in Table~\ref{tab:parameter_set}, assuming the co\nobreakdash-located scenario with a latency of 2 \si{\us}.
In this setting, the combined classical\nobreakdash-side and quantum\nobreakdash-side computation time is about 51 milliseconds,
with an overhead of about 10.5 milliseconds (values below 0.1 milliseconds were deemed negligible).
Thus, the total execution time per iteration, including computation and overhead, is approximately 61.5 milliseconds.
Taking a conservative approach based on the above analysis~\cite{ravi2021quantum}, it is reasonable to assume a queueing time of one minute per iteration.
Under this assumption, the queueing time alone would make up about 99.9\% of the total runtime.
Moreover, even with Pattern 2 defined in Table~\ref{tab:parameter_set}, which entails a longer quantum circuit execution time, queueing time would still account for more than 99\% of the total runtime.

Therefore, these results highlight the crucial importance of reducing queueing time for efficient quantum\nobreakdash-classical hybrid job execution.

\subsubsection{Priority-based Scheduling}
One potential approach to efficiently execute hybrid jobs is to assign higher priority to the quantum computing jobs originating from hybrid jobs.
Such priority\nobreakdash-based scheduling mechanisms are already available on quantum computing platforms from IBM, AWS, and Osaka University (see Table~\ref{tab:quantum_services}).
A common feature among these services is that users upload their programs.
These programs are then executed as hybrid jobs, utilizing both conventional and quantum computing resources provided by the platform.
In contrast, the details of whether these hybrid jobs receive higher execution priority or exclusive access to quantum computing resources differ by platform implementation.

\begin{table*}[tbh]
    \centering
    \caption{Comparison of Prioritized Execution and Exclusive Access in Various Quantum Computing Platforms.}
    \label{tab:quantum_services}
    \begin{tabular}{l|l|l}
    \hline
    \textbf{Provider} & \textbf{Platform} & \textbf{Prioritized Execution/Exclusive Access} \\
    \hline
    IBM~\cite{ibm_execution}  & Qiskit Runtime & 
        \begin{tabular}{l}
            Prioritized execution/exclusive access\\
            switched by execution mode
        \end{tabular}\\
    \hline
    AWS~\cite{aws2024} & Amazon Braket Hybrid Jobs & Prioritized execution \\
    \hline
    Osaka University~\cite{mori2024kyocho} & Quantum Computing Cloud Service & Exclusive access \\
    \hline
    \end{tabular}
\end{table*}

As shown in our previous scenario analysis, queueing times in quantum computing job queues can dominate the overall runtime of hybrid jobs.
Consequently, priority\nobreakdash-based scheduling or resource exclusivity can significantly reduce queueing times.
This, in turn, shortens the total runtime of each quantum\nobreakdash-classical hybrid job.
This reduction in total runtime further contributes to minimizing idle periods for conventional computers.
However, it is important to note that prioritizing hybrid jobs inherently shifts queueing delays to other tasks.
For instance, standalone quantum jobs might face extended queueing times,
while quantum\nobreakdash-classical hybrid jobs could also encounter increased waiting periods in the quantum\nobreakdash-classical hybrid job queue before execution begins.
Thus, relying solely on priority\nobreakdash-based scheduling does not necessarily guarantee an improvement in overall platform throughput.

\subsubsection{Parallel Execution Strategy}
Parallel execution of multiple quantum computing jobs is considered a promising approach to reducing average queueing times and enhancing overall platform throughput~\cite{das2019case}.

Osaka University's quantum computing cloud service recently introduced the feature called ``quantum multiprogramming''~\cite{mori2024heiretsu}.
This allows parallel execution by merging multiple user\nobreakdash-specified quantum programs into a single executable quantum computing job
and subsequently partitioning the results back into the original individual tasks.
Currently, users must explicitly designate which programs to execute in parallel.
However, future developments aim to provide an automated feature that dynamically assesses quantum processor availability and schedules jobs accordingly.

We review two empirical studies that have demonstrated the effectiveness of parallel execution in shortening total quantum computation time.
Reference~\cite{romero2024quantum} conducted an experimental evaluation using the 127\nobreakdash-qubit IBM ``Eagle'' quantum processor.
The study showed that parallel execution of benchmark quantum circuits reduced the total execution time
from approximately 70 seconds (serial execution) to 24 seconds (parallel execution).
This corresponds to a reduction of about 66\%.
Although the measurement distributions differed slightly between serial and parallel executions,
these discrepancies did not significantly impact the final computational results.

Similarly, Reference~\cite{wu2024reducing} conducted a job scheduling experiment using the 66\nobreakdash-qubit Xiaohong quantum processor.
Compared to traditional FIFO scheduling, the implementation of job reordering and parallel execution scheduling
reduced the turnaround time from job submission to completion by about 93\% on average.
Although the scheduling strategies resulted in an about 5\% reduction in the probability of obtaining the desired results,
the authors used strategic qubit mapping to mitigate the quantum noise more effectively than other comparative scheduling approaches.

As the number of qubits in quantum hardware continues to grow, not all quantum computing jobs will require the full qubit capacity.
This scenario naturally creates more opportunities for parallel execution.
In combination with advanced job scheduling strategies, parallel execution thus emerges as a promising capability.
It has the potential to enhance quantum computing job throughput and effectively reduce queueing times.

\subsubsection{Conclusion}
We have examined the hypothesis that appropriate job scheduling strategies can effectively mitigate queueing time issues inherent in quantum\nobreakdash-classical hybrid jobs.

A detailed analysis of job execution data and our scenario\nobreakdash-based analysis indicate that queueing time can dominate the overall execution time of each quantum\nobreakdash-classical hybrid job.
We observed that several current quantum computing cloud services provide prioritized execution or exclusive access for hybrid jobs.
This priority\nobreakdash-based scheduling methods significantly reduce queueing times, thereby shortening the total runtime of each hybrid job.
Moreover, this reduction in total runtime contributes to minimizing idle periods for conventional computers.
However, priority\nobreakdash-based scheduling does not necessarily improve the overall platform throughput.

In contrast, the parallel execution strategy is a promising approach to reducing average queueing times and enhancing overall platform throughput.
We reviewed two empirical studies that demonstrated considerable reductions in total quantum computation time through parallelization.

In conclusion, advanced job scheduling strategies can effectively mitigate queueing\nobreakdash-related inefficiencies.
Examples include priority\nobreakdash-based scheduling for sequential quantum tasks within hybrid jobs and parallel execution of quantum computing jobs.
These improvements are expected to minimize idle periods for conventional computers and enhance the overall throughput of quantum\nobreakdash-classical hybrid computing platforms.

\section{Leveraging HPC for Quantum-Classical Hybrid Computing}
\label{sec:hpc-resource}

In this section, we address the second research question posed in the Introduction:
whether leveraging a full\nobreakdash-fledged HPC system, rather than a standard conventional computer, results in additional benefits.
Specifically, we discuss the computational advantages gained by integrating quantum computers with HPC systems.
We hypothesize that quantum\nobreakdash-classical hybrid algorithms require increasingly substantial conventional computing resources as they are applied to large\nobreakdash-scale, real\nobreakdash-world problems.
In that case, HPC resources can be critical for handling the substantial classical\nobreakdash-side workload that arises in these hybrid workflows.

In the following subsections, we review the existing literature and recent experimental results to substantiate the hypothesized advantages of leveraging HPC resources alongside quantum computing.
In Subsection~\ref{subsec:hybrid_algorithm}, we first analyze the benefits HPC brings to the execution of quantum\nobreakdash-classical hybrid algorithms in fields such as quantum chemistry and quantum machine learning.
Next, in Subsection~\ref{subsec:error_mitigation}, we examine how HPC resources can mitigate the classical\nobreakdash-side overhead inherent in error mitigation techniques.
Finally, in Subsection~\ref{subsec:circuit_partitioning_optimization}, we investigate to what extent HPC capabilities are necessary to address the intensive computations required for quantum circuit partitioning and optimization.

\subsection{Quantum-Classical Hybrid Algorithms}
\label{subsec:hybrid_algorithm}
This subsection investigates the potential benefits of employing HPC resources in quantum\nobreakdash-classical hybrid algorithm execution.
Quantum\nobreakdash-classical hybrid algorithms typically involve alternating computational tasks between quantum and conventional computers.
Representative examples include VQE in quantum chemistry and quantum circuit learning in quantum machine learning.
According to References~\cite{cerezo2021variational, bharti2022noisy}, classical\nobreakdash-side computational roles in these variational algorithms are relatively lightweight,
primarily involving simple gradient calculations and parameter update tasks.

Recently, however, the concept of ``Quantum\nobreakdash-Centric Supercomputing'' has emerged~\cite{schneider2024, alexeev2024quantum}.
It advocates for integrated computational platforms that combine HPC systems with quantum computers to tackle highly complex, real\nobreakdash-world problems.
Given the inherent performance constraints of current NISQ computers, standalone quantum computing solutions can only handle problems of limited complexity.
Therefore, the strategic allocation of significant computational responsibilities to HPC systems and the exploitation of the unique capabilities of quantum computers
offer a promising pathway toward solving practical, large\nobreakdash-scale problems.

The development of integrated quantum\nobreakdash-HPC platforms is in progress, and research on HPC\nobreakdash-supported hybrid algorithms is expected to grow in the near future.
However, practical examples remain scarce at present.
Consequently, our investigation adopts a broader definition of hybrid algorithms,
encompassing not only strict alternation between quantum and classical processing but also the extensive classical\nobreakdash-side computations associated with
pre\nobreakdash-/post\nobreakdash-processing tasks in hybrid job workflows.

Specifically, we surveyed literature across quantum chemistry and quantum machine learning domains to identify cases
involving extensive and complex classical\nobreakdash-side computations exceeding simple gradient calculations.
We identified examples indicating a potential necessity of HPC resources for pre\nobreakdash-processing or post\nobreakdash-processing tasks in hybrid job workflows.
Notably, we found at least one example that explicitly employed HPC resources.

\subsubsection{Case Study: Quantum Chemistry}
In quantum chemistry, Quantum\nobreakdash-Selected Configuration Interaction (QSCI)~\cite{kanno2023quantum} serves as a prime example of a hybrid algorithm
in which classical\nobreakdash-side computing plays a significant role.
QSCI is a novel hybrid algorithm designed for calculating ground\nobreakdash-state and excited\nobreakdash-state molecular energies.
The algorithm leverages quantum computers to sample important electron configurations.
It then restricts the Hamiltonian to a subspace composed of the most frequently sampled configurations.
Subsequently, conventional computers perform the diagonalization of this reduced Hamiltonian.

The benefits of QSCI include improved accuracy over methods such as VQE.
Additionally, it maintains a desirable variational characteristic, ensuring that the calculated energy never falls below the true ground\nobreakdash-state energy,
even in the presence of quantum noise or statistical errors.

As the molecular size or the desired accuracy increases, the Hamiltonian matrix dimension grows, requiring substantial resources from conventional computers.
By harnessing HPC as the conventional computing resource, theoretical estimation made in Reference~\cite{kanno2023quantum} suggests that QSCI has the potential to accurately compute molecular ground\nobreakdash-state energies with an error margin of approximately 0.001 Hartree.
This level of performance is projected for molecules such as the chromium dimer and hydrogen chains, which previously posed challenges for conventional quantum chemistry methods.

In fact, practical applications leveraging HPC resources already exist.
A modified version of the QSCI algorithm, incorporating classical post\nobreakdash-processing steps referred to as self\nobreakdash-consistent configuration recovery,
was implemented using the supercomputer Fugaku~\cite{robledo2025chemistry}.
Specifically, the ground\nobreakdash-state electronic configurations and energies of the iron\nobreakdash-sulfur cluster [Fe\(_4\)S\(_4\)(SCH\(_3\))\(_4\)]\(^{-2}\) were estimated.
The classical post\nobreakdash-processing step alone required approximately 1.5 hours of computation using 6400 nodes of the supercomputer Fugaku,
highlighting the critical role of HPC in tackling complex and large\nobreakdash-scale problems.

\subsubsection{Case Study: Quantum Machine Learning}
In quantum machine learning, Reference~\cite{rudolph2023synergistic} suggests a synergy between conventional and quantum computing resources,
through interconvertibility of tensor networks and Parametrized Quantum Circuit (PQC) representations.
Specifically, pre\nobreakdash-training tensor networks on conventional computers can be used to initialize PQC parameters.
The authors experimentally demonstrated that this initialization method mitigated the barren plateau problem, thereby enhancing the efficiency of quantum model training.

In their experiments, classical tensor network pre\nobreakdash-training was successfully applied to quantum circuit Born machines with up to 100 qubits~\cite{rudolph2023synergistic}.
The Bars and Stripes benchmark dataset was used for evaluation.
Experimental results suggested that scaling up to more complex datasets and larger problems required tensor networks with larger bond dimensions.
Consequently, it is expected that greater conventional computing resources become necessary to tackle larger problems.

Although specific computing resources are not specified in Reference~\cite{rudolph2023synergistic}, the authors argue that only moderate conventional computing resources are necessary for their particular scenario.
They suggest that HPC\nobreakdash-level computational power may eventually be required as quantum machine learning models and associated datasets grow larger.
Nevertheless, they emphasize that the increase in required conventional computing resources will be moderate rather than exponential.

\subsubsection{Conclusion}
We have examined the advantages of employing HPC as the conventional computing resources in quantum\nobreakdash-classical hybrid job workflows,
including pre\nobreakdash-processing, algorithms execution, and post\nobreakdash-processing stages.

We identified case studies that enhance quantum chemistry simulation accuracy and accelerate machine learning processes
by leveraging complex classical\nobreakdash-side computations during pre\nobreakdash-processing and post\nobreakdash-processing phases.
We observed that as the problem size increased, more conventional computing resources were required.
In at least one case, HPC resources were actually utilized.

Therefore, the advantages of integrating HPC as conventional computing resources in quantum\nobreakdash-classical hybrid algorithms become evident,
especially when tackling large\nobreakdash-scale, real\nobreakdash-world problems.
As integrated quantum\nobreakdash-HPC platforms continue to emerge, we anticipate that research in this direction will intensify.

\subsection{Quantum Error Mitigation}
\label{subsec:error_mitigation}
This subsection examines the necessity of HPC resources in managing the classical\nobreakdash-side computational overhead associated with error mitigation techniques.
Current quantum computing hardware is in the NISQ era, which is characterized by non\nobreakdash-negligible noise arising during quantum gate operations and qubit measurements.
Error mitigation techniques aim to reduce noise\nobreakdash-induced inaccuracies by post\nobreakdash-processing quantum measurement results on conventional computers,
thereby recovering outcomes closer to the ideal, noise\nobreakdash-free scenario.

Typically, conventional computers handle post\nobreakdash-processing of the measurement data obtained from quantum computers,
and an increase in data volume generally translates into higher computational and memory demands on conventional computing resources.
Therefore, it is crucial to evaluate how the number of qubits, quantum circuit depth, and execution repetitions affect the size of measurement data sets.
Additionally, quantifying the computational complexity and memory requirements associated with processing these measurement results on conventional computers
will help clarify the potential benefits of HPC platforms, such as accelerated processing and enhanced memory capacity.

In the following analysis, we reviewed relevant literature and experimental data to clarify the scalability challenges that error mitigation poses for larger\nobreakdash-scale quantum circuits.
These circuits, characterized by increasing the number of qubits and circuit depths, are expected to become feasible in the near future as hardware advances.

\subsubsection{Conventional Computing Resource Requirements}
As discussed in Subsection~\ref{subsec:high_bandwidth}, increasing the number of qubits and circuit depth leads to an exponential growth in the measurement data required for effective error mitigation~\cite{takagi2023universal,tsubouchi2023universal}.
We denote the size of this measurement data by \(N\), and analyze the computational complexity and memory requirements of classical post\nobreakdash-processing tasks for representative error mitigation methods.

For representative methods of gate operation error mitigation, such as Zero Noise Extrapolation and Probabilistic Error Cancellation,
classical post\nobreakdash-processing primarily involves computing average values of measurement outcomes~\cite{cai2023quantum}.
Consequently, computational complexity and memory requirements both scale as \(O(N)\).

A representative measurement error mitigation technique, known as Matrix\nobreakdash-free Measurement Mitigation (M3),
relies on solving linear equations using Krylov subspace methods~\cite{nation2021scalable}.
In the worst case, for a sufficiently large number of qubits, these equations can form \(N \times N\) systems.
This leads to computational complexity and memory requirements that both scale as \(O(N^2)\).

Since the computational complexity and memory requirements for these representative techniques range from \(O(N)\) to \(O(N^2)\),
once \(N\) grows exponentially with the size of quantum circuits,
the conventional computing resources required will likewise grow at an exponential rate.

\subsubsection{Case Study: Computing Resources for Error Mitigation}
Having established the general trend in conventional computing resource demands for error mitigation,
we now investigate a specific case study to better understand the current scale of resources required.
We reviewed the existing literature that explicitly measures conventional computing resource usage,
such as processing time and hardware specifications.

A notable study~\cite{pokharel2024scalable} conducted comparative experiments between M3 and a proposed method referred to as Iterative Bayesian Unfolding (IBU).
Both methods were tested on benchmark GHZ\nobreakdash-state quantum circuits involving up to 127 qubits, executed 100,000 times.
The classical\nobreakdash-side computations required for error mitigation were carried out using a conventional computer equipped with an NVIDIA GeForce RTX 3090 GPU.

The experimental result at the 81 qubits demonstrated processing times of approximately 50 seconds for M3 and under 200 seconds for IBU.
However, experiments involving circuits larger than 81 qubits failed entirely for both methods,
indicating that the probability of obtaining ideal noise\nobreakdash-free measurement results became zero after mitigation.
Even below 81 qubits, the discrepancy between ideal and mitigated probability distributions increased significantly as the number of qubits rose.

General theory on measurement error mitigation techniques indicates that achieving a given error mitigation accuracy requires the number of shots to grow exponentially with the number of qubits~\cite{bravyi2021mitigating}.
In the above study, the number of shots remained fixed at 100,000.
This could contribute to the observed rapid degradation in accuracy with increasing the number of qubits.
Although this study did not employ HPC\nobreakdash-level resources, it suggests that achieving higher accuracy would necessitate a significantly larger number of shots and,
consequently, larger conventional computing resources.

However, the critical question arises: would simply increasing conventional computing resources solve this issue?
With the number of shots exceeding 100,000 and growing exponentially, not only would the quantum computation time increase,
but the dimensionality of the linear equations solved by conventional computers would also expand exponentially,
potentially overwhelming both quantum and conventional computing resources.
Therefore, addressing this problem will likely require reducing the qubit measurement error rate in addition to scaling conventional computing resources.
As discussed in the Appendix, it is theoretically feasible to maintain error mitigation accuracy and keep the number of shots constant 
despite scaling the number of qubits by a factor of \(k\) if the error rate is reduced to \(1/k\).

According to Reference~\cite{pokharel2024scalable}, the quantum computer utilized in their experiments, IBM Washington (Eagle r1, 127\nobreakdash-qubit),
exhibited an average measurement error rate of approximately 3.51\%.
In their experiments involving 10 qubits with 100,000 shots,
the accuracy metric reached about 0.8, where the metric was measured using the \(\ell_1\)\nobreakdash-norm between mitigated and ideal probability distributions (normalized as a score of 1 indicating perfect agreement).
Extrapolating from these findings, reducing the average measurement error rate to 0.351\% could potentially allow processing up to 100 qubits at the same accuracy and the number of shots.

\subsubsection{Conclusion}
We have assessed the necessity of HPC resources for classical post\nobreakdash-processing in error mitigation techniques.

In a case study, post\nobreakdash-processing was performed on single GPU, which proved sufficient for the current workloads without incurring prohibitively long processing times.
This computational load did not warrant the use of HPC resources.

Nevertheless, our theoretical analysis suggests exponential growth in conventional computational requirements as quantum circuits become deeper and involve more qubits.
Thus, future scalability in quantum computing may inevitably benefit from integrating HPC for classical post\nobreakdash-processing tasks associated with error mitigation.
We also note that, even in the aforementioned case study, if higher error mitigation accuracy is desired, significantly larger conventional computing resource could be required.

It is also important to acknowledge that solely increasing computing resources and the number of shots may not offer a viable long\nobreakdash-term solution,
since this approach could rapidly exhaust the capabilities on both quantum and conventional computing resources.
Consequently, improvements in quantum hardware error rates must accompany scaling efforts to effectively manage the computational burden of quantum error mitigation.

\subsection{Quantum Circuit Partitioning and Optimization}
\label{subsec:circuit_partitioning_optimization}

This subsection investigates the necessity of HPC resources in managing the classical\nobreakdash-side computational overhead inherent in quantum circuit partitioning and optimization procedures.

\paragraph*{Quantum Circuit Partitioning}
Due to the limited number of qubits in current quantum hardware, large\nobreakdash-scale quantum circuits cannot be directly executed.
Furthermore, as circuit size increases, quantum computational errors accumulate with circuit depth, leading to a deterioration of execution fidelity.
To address these challenges, various techniques have been developed to partition large quantum circuits into smaller,
more manageable sub\nobreakdash-circuits that fit within the available hardware constraints~\cite{bravyi2016trading, peng2020simulating}.
To reconstruct the original circuit output from the measurement results of the partitioned sub\nobreakdash-circuits, computationally intensive classical post\nobreakdash-processing tasks are performed.
Finding an optimal partitioning that minimizes the cost of classical post\nobreakdash-processing can be formulated as a combinatorial optimization problem.
These optimization problems constitute classical pre\nobreakdash-processing tasks.

\paragraph*{Quantum Circuit Optimization}
Executing quantum circuits on actual quantum hardware requires compiling them into an executable form that is physically realizable on the target quantum computer,
while still maintaining logical equivalence with the original circuit~\cite{ge2024quantum}.
Effective compilation involves optimizing the quantum circuit structure to minimize gate count and circuit depth,
which reduces the quantum error rate and improves overall execution fidelity.
Two\nobreakdash-qubit gates, in particular, exhibit higher error rates than single\nobreakdash-qubit gates; hence, optimization strategies frequently focus on reducing the number of two\nobreakdash-qubit operations.
Such optimization tasks potentially necessitate significant conventional computing resources.

Given these two critical aspects, partitioning and optimization, we hypothesize that HPC resources will offer tangible performance advantages
for dealing with the classical pre\nobreakdash-/post\nobreakdash-processing overhead, which increases significantly with the size and complexity of quantum circuits.
Consequently, to test this hypothesis we explored existing empirical cases where quantum circuit partitioning and optimization tasks were executed.
In particular, we focus on their computational runtimes and the specifications of the conventional hardware used.

\subsubsection{Case Study: Circuit Partitioning}
We examined empirical studies of quantum circuit partitioning, focusing on two prominent techniques: quantum circuit cutting and quantum circuit knitting.

Quantum circuit cutting is based on a so\nobreakdash-called wire\nobreakdash-cut technique, where quantum circuits are partitioned by cutting edges between gates,
resulting in multiple sub\nobreakdash-circuits with fewer qubits.
However, reconstructing the output of the original circuit from measurements of these sub\nobreakdash-circuits incurs a computational cost on the classical side,
scaling exponentially with the number of cuts~\cite{peng2020simulating}.

An empirical example of quantum circuit cutting is provided by CutQC~\cite{tang2021cutqc, tang2022cutting}.
The authors demonstrated the cutting of benchmark circuits ranging from 40 to 100 qubits.
The benchmark circuits were partitioned into sub\nobreakdash-circuits, each comprising up to 75\% of the original number of qubits.
The partitioning step was formulated as a mixed\nobreakdash-integer programming (MIP) problem solved efficiently using the Gurobi solver.
In their experiment, the MIP problems were solved within minutes even for 100\nobreakdash-qubits circuits.
Thus, the pre\nobreakdash-processing runtimes were considered negligible and no further details were given.
However, the classical post\nobreakdash-processing step to reconstruct the outputs of the original circuits from sub\nobreakdash-circuits measurements was computationally intensive.
The most computationally intensive case was the cutting of 100\nobreakdash-qubit quantum supremacy circuit~\cite{boixo2018characterizing},
which served as the basis for Google's quantum supremacy experiment~\cite{arute2019quantum}.
In this case, the runtime was over \(10^3\) seconds using an NVIDIA A100 GPU.

Quantum circuit knitting extends the cutting methodology by introducing a gate\nobreakdash-cut technique,
which decomposes two\nobreakdash-qubit gates into multiple single\nobreakdash-qubit gates.
Similar to cutting, reconstructing the original circuit output from sub\nobreakdash-circuits also results in classical computational costs that scale exponentially with the number of cuts~\cite{piveteau2023circuit}.

An empirical example of quantum circuit knitting is presented in Reference~\cite{ren2024hardware}.
The authors demonstrated the knitting of benchmark circuits ranging from 10 to 80 qubits.
The knitting wes executed in the scenario where these circuits exceeded the available quantum hardware capacity by a factor of 1.5 to 4.
Partitioning was performed using the Karger\nobreakdash-Stein algorithm, which was claimed to potentially offer better scalability compared to the MIP solver employed by CutQC.
Thus, the pre\nobreakdash-processing runtimes were also ignored as in CutQC.
However, the classical post\nobreakdash-processing remained computationally intensive.
The most computationally intensive case was the knitting of 50\nobreakdash-qubit quantum supremacy circuit~\cite{boixo2018characterizing}.
In this case, the runtime was approximately \(10^4\) seconds using an NVIDIA V100 GPU.

These two experimental cases showed that classical post\nobreakdash-processing runtimes on a single GPU reached to the order of \(10^3\) to \(10^4\) seconds.
Although HPC\nobreakdash-level resources were not employed in these studies, they could become beneficial if computational requirements grow exponentially in the future.
However, we also note that the number of available qubits continues to increase, which could reduce the need for circuit partitioning.

\subsubsection{Case Study: Circuit Optimization}
When executing quantum circuits on quantum hardware, a large number of error\nobreakdash-prone two\nobreakdash-qubit gates(e.g., CNOT gates) can degrade overall computational fidelity.
To mitigate this degradation, quantum circuit optimization techniques aim to minimize the number of two\nobreakdash-qubit gates while preserving the logical equivalence.

Reference~\cite{schneider2023sat} presents an optimization method specifically tailored for Clifford circuits, a subclass of quantum circuits.
This method transforms the computation of a quantum circuit into a stabilizer tableau representation,
and then formulates a combinatorial optimization problem to find a quantum circuit implementing this representation with the minimal number of two\nobreakdash-qubit gates.
This optimization problem is reduced to a Boolean satisfiability (SAT) problem.
Solving SAT problems typically requires conventional computing resources that grow exponentially with the number of qubits.
In their experiments, the authors optimized random Clifford circuits up to 26 qubits on a conventional computer equipped with an AMD Ryzen 9 5950X CPU (16\nobreakdash-core, 5 GHz) and 128 GiB memory.
For a 26\nobreakdash-qubit circuit, the optimization required on the order of \(10^4\) seconds.
In this experiment, heuristic methods~\cite{aaronson2004improved, bravyi2021clifford} were also tested for comparison.
While these heuristic methods significantly reduced computational time, the resulting circuits contained more two\nobreakdash-qubit gates compared to those found by optimal approach.

Although optimizing general quantum circuits beyond Clifford circuits remains a challenging open problem,
subsequent research~\cite{peham2023depth} has extended the approach to circuits composed of the universal Clifford+T gate set,
albeit currently focusing only on Clifford sub\nobreakdash-circuits.
We anticipate that general circuit optimization will require even greater conventional computing resources.

In contrast to circuit partitioning, as the size of quantum circuits continues to grow,
the computational hardness of circuit optimization increases substantially.
Consequently, HPC\nobreakdash-level resources may be beneficial in the future.

\subsubsection{Conclusion}
We have evaluated the potential necessity of HPC resources for classical processing tasks in quantum circuit partitioning and optimization.

In quantum circuit partitioning, we primarily focused on techniques such as cutting and knitting.
Both techniques necessitate resource\nobreakdash-intensive classical post\nobreakdash-processing tasks whose computational complexity scales exponentially with the number of cuts.
Experimental cases showed that classical post\nobreakdash-processing runtimes on a single GPU reached approximately \(10^3\) seconds for cutting a 100\nobreakdash-qubit circuit and about \(10^4\) seconds for knitting a 50\nobreakdash-qubit circuit.
If the number of cuts increases in the future, employing HPC resources for classical processing will become increasingly beneficial.
However, given the anticipated growth in the available number of qubits, it remains unclear whether the required number of cuts will continue to increase.

In quantum circuit optimization, we reviewed a SAT\nobreakdash-based approach for minimizing two\nobreakdash-qubit gates.
This technique also exhibits computational complexity that grows exponentially with increasing circuit size.
One study demonstrated that optimizing a 26\nobreakdash-qubit Clifford circuit required approximately \(10^4\) seconds using a single 16\nobreakdash-core CPU.
General quantum circuit optimization beyond Clifford circuits is expected to involve even higher computational costs.
Given the anticipated growth in the size of quantum circuits, employing HPC resources could become increasingly advantageous.

\section{Summary, Conclusion and Future Directions}
\label{sec:conclusion}

In this paper, we have conducted a systematic survey on the advantages of quantum\nobreakdash-HPC platforms.
Given the current constraints and limitations inherent to NISQ computers, we examined the feasibility and potential benefits of
co\nobreakdash-locating quantum computers with HPC systems connected via high\nobreakdash-speed networks, particularly targeting near\nobreakdash-future industrial use cases such as quantum chemistry and quantum machine learning.
We briefly summarize the results of our survey and outline future research directions.

\subsection{Summary and Conclusion}
\label{subsec:summary_conclusion}
The first research question posed in the Introduction is whether the co\nobreakdash-location of quantum and conventional computers connected via high\nobreakdash-speed networks provides tangible advantages.
To address this question, we summarize our key findings on the benefits derived from co\nobreakdash-locating quantum and conventional computers.
Regarding latency reduction (see Subsection \ref{subsec:latency_reduction}), our model\nobreakdash-based analysis demonstrated that a co\nobreakdash-located configuration improves job throughput compared to a remotely located configuration.
However, the extent of this benefit is reduced under certain conditions, specifically when quantum circuit execution times become longer or when the number of shots per iteration increases.
Nevertheless, the advantage still exists when evaluating overall throughput for diverse workloads on the hybrid computing platform.

Regarding increased bandwidth (see Subsection \ref{subsec:high_bandwidth}), our investigation highlighted that quantum error mitigation techniques in the NISQ era
lead to an exponential growth in measurement data transmitted from quantum to conventional computers, particularly as one or both of the number of qubits and the circuit depth increase.
Although current data volumes remain relatively modest, our model\nobreakdash-based analysis and the projected growth in data volume suggest that
establishing high\nobreakdash-bandwidth communication through co\nobreakdash-location of quantum and conventional computers will be essential to maintain high throughput in the near future.

Regarding advanced job scheduling (see Subsection \ref{subsec:job_scheduling}), our analysis showed that queueing times significantly impact quantum\nobreakdash-classical hybrid job runtimes.
We found that advanced scheduling strategies, such as prioritizing sequential quantum computing jobs within hybrid jobs and employing parallel execution of quantum computing jobs, effectively reduce queueing times.
These advanced scheduling strategies are critical for minimizing idle periods for conventional computers and improving overall throughput of hybrid jobs.

The second research question is whether leveraging a full\nobreakdash-fledged HPC system, rather than a standard conventional computer, results in additional benefits. 
To address this question, we summarize our key findings on the potential benefits of employing HPC as the conventional computing resource integrated with quantum computing platforms.
For hybrid algorithm execution (see Subsection \ref{subsec:hybrid_algorithm}), we identified clear advantages of integrating HPC resources, particularly in pre\nobreakdash-processing and post\nobreakdash-processing phases.
Case studies in quantum chemistry and quantum machine learning indicate that extensive classical\nobreakdash-side computations are vital to enhancing algorithm accuracy and performance, especially as problem scales increase.
Consequently, integrating HPC capabilities into quantum\nobreakdash-classical computational frameworks is essential for effectively addressing large\nobreakdash-scale, real\nobreakdash-world problems.

In terms of error mitigation (see Subsection \ref{subsec:error_mitigation}), our analysis showed that while current error mitigation tasks are typically manageable
using conventional GPU resources without necessitating HPC\nobreakdash-level computational power, theoretical analyses indicate exponential growth in classical\nobreakdash-side computational requirements with increasing quantum circuit depth and the number of qubits.
Therefore, as quantum circuits scale, there is potential to utilize HPC resources for quantum error mitigation.
However, solely relying on increased conventional computing resources is unsustainable in the long term, underscoring the importance of improving error rates of quantum hardware alongside computational capacity expansion.

Concerning quantum circuit partitioning and optimization (see Subsection \ref{subsec:circuit_partitioning_optimization}), our survey revealed substantial computational demands associated with classical post\nobreakdash-processing, particularly quantum circuit cutting and knitting, and pre\nobreakdash-processing tasks such as circuit optimization.
Indeed, current implementations primarily rely on multiprocessor CPUs and GPUs, with computational requirements already extending to hours for moderate\nobreakdash-sized circuits.
While the necessity for HPC resources in quantum circuit partitioning remains uncertain due to potentially reduced demand as the available number of qubits increases,
quantum circuit optimization clearly exhibits exponential computational demands as circuit sizes grow.
Therefore, integrating HPC resources offers potential advantages for efficiently handling larger optimization tasks in the future.

Finally, we provide overall conclusions from our analysis.
Through this survey, we identified evidence from existing literature for the advantages of quantum\nobreakdash-HPC platforms across multiple aspects aligned with the research questions posed in the Introduction.
Although isolated studies addressing specific aspects exist, comprehensive evaluations from an integrated, holistic perspective are still lacking.
The platform\nobreakdash-specific requirements for communication, job scheduling, and conventional computing resources ideally should be tailored comprehensively to particular use cases.
However, such in\nobreakdash-depth analyses have not yet been widely conducted.

\subsection{Future Directions}
The development of integrated quantum\nobreakdash-HPC platforms is actively underway at various locations worldwide, offering promising pathways for advancing computational capabilities.
Future research is anticipated to conduct comprehensive evaluations from the integrated, holistic perspective described in Subsection~\ref{subsec:summary_conclusion}.
Specifically, such evaluations should focus on error mitigation, quantum circuit partitioning and optimization, and throughput enhancement, as these are essential components
contributing to improved operational capabilities and performance of quantum\nobreakdash-HPC platforms.
Beyond individual technological improvements, studies that evaluate overall platform performance and the broader impact of specific technical enhancements are also crucial.

Enhancements to usability are highly desirable, where users are freed from the responsibility of manually implementing individual technological components.
Instead, platforms should autonomously manage these aspects to optimize job execution effectively.

From the perspective of use cases, there is a strong expectation for algorithmic developments capable of addressing problems at real\nobreakdash-world application scales.
Following the emerging concept of ``Quantum\nobreakdash-Centric Supercomputing,'' research into hybrid algorithms assigning increased roles to conventional computing resources is in its early stages,
and novel algorithmic propositions are expected to emerge continually.
By leveraging the unique strengths of both NISQ computers and HPC systems, it is anticipated that entirely new application domains will become feasible, extending the boundaries of computational science.

Overall, the integration of quantum\nobreakdash-HPC platforms presents a compelling vision for accelerating scientific discovery.
As NISQ computers gradually mature, the synergy between quantum computers and HPC systems is expected to unlock novel computational regimes
and provide a pathway toward practical, large\nobreakdash-scale quantum\nobreakdash-classical hybrid computation.

\vspace{1em}

\begin{acknowledgments}
    This work was supported by the New Energy and Industrial Technology Development Organization (NEDO) under the project JPNP20017, titled ``Research and Development Project of the Enhanced Infrastructures for Post\nobreakdash-5G Information and Communication Systems.''
\end{acknowledgments}

\appendix*

\section{Relationships Among Error Rates, Number of Qubits and Required Number of Shots for Error Mitigation}
\label{appendix:measurement_error}

In this appendix, we discuss the relationship among measurement error rates, the number of qubits and the required number of shots for effective error mitigation.

We begin by referring to a theoretical analysis~\cite{bravyi2021mitigating}, which provides a foundational formula for estimating the required number of shots.
Let \(M\) be the number of shots needed to ensure that the mitigated expectation value deviates from the ideal noiseless expectation value
by no more than \(\delta\), with probability at least \(2/3\).
This required number of shots can be expressed as:
\begin{equation}
    M = 4\delta^{-2}\Gamma^2,
\end{equation}
where \(\Gamma^2\) referred to as the ``error mitigation overhead'' in the reference.
Considering an \(N\)\nobreakdash-qubit quantum circuit, let \(\epsilon_j\) and \(\eta_j\) denote the measurement error rates for the \(j\)\nobreakdash-th qubit,
specifically the probabilities of incorrectly measuring a \(\ket{0}\) state as \(\ket{1}\) and a \(\ket{1}\) state as \(\ket{0}\), respectively.
Under the assumption that \(\epsilon_j, \eta_j \ll 1\), \(\Gamma^2\) can be approximated as:
\begin{equation}
    \Gamma^2 \approx e^{4\gamma},\ \gamma=\sum^N_{j=1}\max\{\epsilon_j, \eta_j\}.
\end{equation}

Subsequently, leveraging the theoretical relationships described above, we derive an approximate expression that provides deeper insight into the interplay among measurement error rates, the number of qubits,
and the required number of shots for error mitigation.
Assuming \(\epsilon_j \approx \eta_j\) for all qubits, we define the average measurement error rate as \(\bar{\epsilon}\approx \sum^N_{j=1}\epsilon_j / N\).
Under this assumption, \(\gamma\) can be approximated by \(\gamma \approx N\bar{\epsilon}\).
Thus, the required number of shots simplifies to:
\begin{equation}
    M \approx 4\delta^{-2} e^{4 N \bar{\epsilon}}.
\end{equation}
A notable observation from this approximation is that \(N\) (the number of qubits) and \(\bar{\epsilon}\) (the average measurement error rate) appear as a product in the exponential.
Consequently, if the number of qubits increases by a factor of \(k\), one can maintain the same number of shots (and the same computational accuracy)
by reducing the average measurement error rate by a factor of \(1/k\).
This result underscores the critical importance of reducing qubit measurement error rates to efficiently manage computing resources as the number of qubits scale.

\bibliography{bibliography}

\end{document}